\documentclass[
 aps,
 pre,
 amsmath,amssymb,
 reprint,
showkeys
]{revtex4-2}

\usepackage[T1]{fontenc}
\usepackage[utf8]{inputenc}
\usepackage{amsmath}
\usepackage{amssymb}
\usepackage{graphicx}
\usepackage{refstyle}
\usepackage{skak}
\usepackage{multirow}
\usepackage{epsdice}
\usepackage{soul}
\usepackage[
  citecolor=blue,
  colorlinks,
  linkcolor=blue,
  urlcolor=blue,
]{hyperref}
\usepackage{epstopdf}
\usepackage{bm}
\usepackage[version=4]{mhchem}
\usepackage{tikz}
\usetikzlibrary{arrows,shapes,trees}
\usepackage{lipsum} 

\begin{document}
	\title{Hostility prevents the tragedy of the commons in metapopulation with asymmetric migration: A lesson from queenless ants}

\author{Joy Das Bairagya}
\email{joydas@iitk.ac.in}
\affiliation{
  Department of Physics,
	Indian Institute of Technology Kanpur, Uttar Pradesh, PIN: 208016, India
}
\author{Sagar Chakraborty}
\email{sagarc@iitk.ac.in}
\affiliation{
  Department of Physics,
	Indian Institute of Technology Kanpur, Uttar Pradesh, PIN: 208016, India
}
\begin{abstract}
{\color{black} A colony of the queenless ant species, \emph{Pristomyrmex punctatus}, can broadly be seen as consisting of small-body sized worker ants and relatively larger body-sized cheater ants. Hence, in the presence of inter-colony migration, a set of constituent colonies act as a metapopulation exclusively composed of cooperators and defectors. Such a set-up facilitates an evolutionary game-theoretic replication-selection model of population dynamics of the ants in a metapopulation. Using the model, we analytically probe the effects of territoriality induced hostility. Such hostility in the ant-metapopulation proves to be crucial in preventing the tragedy of the commons, specifically, the workforce, a social good formed by cooperation. This mechanism applies to any metapopulation---not necessarily the ants---composed of cooperators and defectors where inter-population migration occurs asymmetrically, i.e., cooperators and defectors migrate at different rates. Furthermore, our model validates that there is evolutionary benefit behind the queenless ants' behavior of showing more hostility towards the immigrants from nearby colonies than those from the far-off ones. In order to  calibrate our model's parameters, we have extensively used the data available on the queenless ant species, \emph{Pristomyrmex punctatus}.

}
\end{abstract}

\keywords{Evolutionary Games, Prisoner's Dilemma, Public Goods Game, Cooperation, Tragedy of the Commons, Migration, Ants.}
\maketitle
\section{Introduction}
Territoriality in a metapopulation~\cite{Bay2008,Dunham1999,Fortuna2006,Ojanen2013,Hanski1998,2004Ecol,Veraart2011,Dai2013,Scheffer2009}---a set of spatially separated populations weakly interacting through migration---is manifested as hostile behavior by the individuals, e.g., 
bacteria~\cite{Beer2009,Eijsink2002,Czrn2002}, insects~\cite{Baker1983,Adams1990,Tanner2006}, birds~\cite{Tinbergen1957,Hinde2008,Woolfenden1977,Carlson1986}, carnivores~\cite{Mosser2009,Furrer2011}, and primates~\cite{Kitchen2004,smuts2008primate}. This hostility is often fatal, e.g., in eusocial insects~\cite{Cassill2008,TSUJI1986}, for a fraction of immigrants whose movement between populations are very important to prevent the populations, or even the entire metapopuation, from going extinct: Often, inadequate cooperation among the individuals in a completely isolated population brings about its tragic extinction~\cite{ouborg1993isolation}.

Any extinction effectively occurs because of over-exploitation of resources needed for sustenance. In any population---for simplicity, assumed to be composed of only cooperators and defectors---social goods may be created by cooperation~\cite{Rankin2007}: Evolution leading to the dearth of cooperators in a population would lead to a tragedy of the commons (TOC)~\cite{hardin1968commons}, i.e., the extinction of the social good, and consequently, the population would extinct. Thus, the fact that we witness stable populations all around us means that some level of cooperation is being maintained everywhere.

But how cooperation, or its stricter form---altruism~\cite{Krebs1970}, emerges and is sustained within the theory of evolution in unicellular organisms to complex human societies~\cite{smith1997major} is an enigmatic multi-faceted conundrum: On the face of it, cooperation seems to evolve contradicting the common wisdom of the natural selection~\cite{Chiappin1999,Szolnoki2018,Wu2019,Roca2009,Fu2008,Santos2016,perc2016,Ginsberg2018,Du2018,Szolnoki2013}. Careful analyses, nevertheless, divulge several mechanisms of establishing cooperation~\cite{Nowak2006}, e.g., kin selection~\cite{grafen1985oxford,Taylor1992,Queller1992,FRANK2019,West2002,FOSTER2006}, direct reciprocity~\cite{Hilbe2018}, indirect reciprocity~\cite{Nowak2005}, network reciprocity~\cite{NOWAK2006Book} and group selection~\cite{WYNNEEDWARDS1963,Wilson1975,Taylor1988,Rogers1990,michod1999darwinian,Paulsson2002,Rainey2003,Wilson2005,Traulsen2006}.

The aforementioned analyses are best systematically mathematized through the formalism of the evolutionary game theory~\cite{weitz2016oscillating, Szolnoki2014,Wu2017,Axelrod1981,Traulsen2006,Gerlee2019} using the paradigmatic games of prisoners' dilemma (PD)~\cite{Perc2008,Fu2009,Mobilia2012} and public goods (PGG)~\cite{Du2014,Perc2011}. In this paper, we use this paradigm to assert that hostility towards immigrants in metapopulation is essential for not only preventing the TOC but also to sustain the metapopulation at its maximum potential. In this context, we recall that whether the TOC means complete or partial destruction of the resource leads to the nomenclature of collapsing or component TOC's respectively. In this language, this paper shows that rather intuitive positive effect of asymmetric migration (unequal migration rates of cooperators and defectors)~\cite{Wakano2009,Limdi2018} in proliferating cooperation is not enough to avert component TOC which, however, is tamed by the hostility. 

Hostile behavior towards intruders depends on whether they are neighbors or they belong to a distant territory. Some animal species show less hostile behavior towards the neighbors than the distant ones~\cite{Wilson1975,Eisenberg1988}. This behavior is termed as dear-enemy effect which is evolutionarily beneficial since cost of escalated contests~\cite{Wilson1975} with repeatedly encountered neighbors is avoided.  Whereas, some eusocial insects show the opposite behavior: They are more hostile towards the neighbors than the distant ones~\cite{Gordon1989,Dunn1999,sanada2003encounter}. This behavior is termed as nasty-neighbor effect.

While the perspective presented above is omnipresent, it suffices for the purpose of this paper to consider the illustrative example of the metapopulation of the ant species, {\it Pristomyrmex punctatus}, where the ToC~\cite{Dobata2013}, asymmetric migration~\cite{Dobata2010}  and hostility~\cite{TSUJI1986,Tsuji1988,sanada2003encounter} are well-documented. Arguably, this ant society is a real-world representative of the similar systems across the living world and hence any insight gained from its analysis is in general qualitatively applicable elsewhere as well.

The society of this ant is very intriguing: The queen caste is not present in these ants' colony. All workers are parthenogenetically developed from unfertilized eggs laid by virgin female workers~\cite{Itow1984}. The division of labor in the smaller body sized ants is also unique: All young workers remain inside the nest and participate in reproduction and brood-caring, and old workers become sterile and come out of the nest for external work~\cite{Tsuji1990}. All ants of this species reproduce only once in their lives, in summer~\cite{Itow1984,Tsuji1988a}.

Interestingly, it is very convenient to model this ant system as consisting of only cooperators and defectors:
Some larger body-sized ants (L-type) of the same species are also present in the colony with the smaller ant workers (S-type), but they do not participate in any work except reproduction and always remain inside the nest~\cite{Itow1984}. Accordingly, the L-type workers are the cheater or defector lineage, and S-type workers are the cooperators~\cite{Dobata2008}. The fitness of a cheater is always higher than that of a worker~\cite{Dobata2013,Tsuji1995}, which could lead the population to extinction~\cite{Dobata2010,Tsuji1995}. Thus, the existence of this ant colony itself is a mystery.

Of course, rarely does a colony exist in complete isolation. A metapopulation of the colonies with asymmetric migration between them is observed. The migration is such that the defectors migrate at a higher rate than the cooperators~\cite{Dobata2010}.  Furthermore, the ants are quite hostile towards the immigrant ants which are not from the same natal colony~\cite{TSUJI1986,sanada2003encounter}. {\color{black} They show more hostility towards the neighboring colonies than the far-off colonies~\cite{sanada2003encounter}, i.e., nasty-neighbor effect.} As we argue in this paper, in the backdrop of asymmetric migration, the role of aggressive hostile behavior is to further strengthen the cooperation inside each colonies and help it proliferate across the entire metapopulation.

The interest in theoretical modeling of this ant system is rather recent. Specifically, Young and Belmontein~\cite{Young2018}, set up an interesting model to investigate the dependence of extinction of colonies on migration. The model has coexistence of cooperators and defectors artificially put into it: In the absence of any migration, an isolated colony doesn't go extinct, even though the literature~\cite{Dobata2010,Dobata2013,Tsuji1995} suggests that the fitness of an isolated colony decreases in the presence of the defectors and eventually it collapses. Furthermore, while careful observations~\cite{Dobata2013} clearly concludes that the surviving probability of the defectors is always higher than the cooperators, hence considering the death rate of the cooperators to be higher than that of the defectors in the aforementioned model~\cite{Young2018} seems not too appropriate for the ants. Finally,  the model~\cite{Young2018} adopts a migration mechanism---migration from lower to higher fitness colony---that is not practically feasible for the ants as it requires prior knowledge about the demographic structure of other colonies: It seems to be a very improbable prospect for an ant to know and process all the information required to selectively migrate towards a higher fitness colony.

In view of the above and the fact that we are after a general description of a large number of metapopulations with qualitatively similar driving forces, let us present a very simple yet nontrivial model within the tenet of natural selection. For concreteness and lucidity of the presentation, we would keep the case of {\it Pristomyrmex punctatus} in mind in what follows. 

\section{Game theoretical model}
Let us consider a metapopulation consisting of $M$ different population of a single asexual species. Every population exhaustively contains two classes of individuals: cooperators and defectors.  Here the cooperation is a socially desirable action, while the defection is solely self-interest driven action. The state of the $i$th population can be defined using two variables---$N_i^C$ and $N_i^D$---respectively the total numbers of cooperators and defectors. Due to the finite carrying capacity of any habitat, the death rate of an individual depends on the total number of individuals in a population; consequently, we consider that the death rate of an individual linearly proportional to the total number of individuals, and the proportionality constant, $\delta$, for the cooperators and the defector to be same. Furthermore, if the fitness of the cooperators and the defectors of the $i$th population be $f_i^C(N_i^C, N_i^D)$ and $f_i^D(N_i^C, N_i^D)$ respectively, then we can write the growth dynamics as follows:
\begin{subequations}
\label{eq:cd}
\begin{eqnarray}
	\dot{N}_i^C&=&N_i^C\left[f_i^C(N_i^C,N_i^D)-\delta(N_i^C+N_i^D)\right],\label{evolution_Cooperator_without_migration}\\
\dot{N}_i^D&=&N_i^D\left[f_i^D(N_i^C,N_i^D)-\delta(N_i^C+N_i^D)\right].\label{evolution_Defector_without_migration}
\end{eqnarray}
\end{subequations}
The fitnesses depend on the details of the strategic interactions between the individuals.

The simplest nontrivial strategic interaction---mathematically, realized through Prisoners' dilemma (PD)---manifests the selfishness of the individuals. Any two-player two-strategy one-shot game can be expressed in a compact bimatrix form as follows:
\begin{eqnarray*}  
	\centering
	\begin{tabular}{cc|c|c|}
		& \multicolumn{1}{c}{} & \multicolumn{2}{c}{{Player $2$}}\\
		& \multicolumn{1}{c}{} & \multicolumn{1}{c}{Cooperate} & \multicolumn{1}{c}{\,\,\,\,Defect\,\,\,\,}\\\cline{3-4} 
		\multirow{2}*{{Player $1$}} & Cooperate & $1,1$ & $S,T$ \\\cline{3-4}
		& Defect & $T,S$ & $0,0$ \\\cline{3-4} 
	\end{tabular}\quad
\end{eqnarray*}
where the first and the second element of each cells is the payoffs of the Player~1 and Player~2, respectively. The aforementioned game describes the PD game if the payoffs maintain the following ordinal relationship: $T>1>0>S$.  Unless otherwise specified, in the rest of this paper, we stick to the above ordinal relationship between the payoffs. The average payoffs, i.e., fitnesses  of a cooperator and  a defector is given by
\begin{subequations}
\begin{eqnarray}
	f_i^C(N_i^C,N_i^D)&=&1\cdot \frac{N_i^C}{N_i^C+N_i^D}+S\cdot  \frac{N_i^D}{N_i^C+N_i^D},\\
	f_i^D(N_i^C,N_i^D)&= &T\cdot\frac{N_i^C}{N_i^C+N_i^D},
\end{eqnarray}
\end{subequations}
respectively.  Since $	f_i^C(N_i^C,N_i^D)<f_i^D(N_i^C,N_i^D)$ $\forall N_i^C,N_i^D$ for PD, it is apparent to see the defectors  thrives in the population. However, when there are no cooperators left in the population, the fitness of a defector becomes zero, i. e., $f_i^D(N_i^C=0,N_i^D)=0$; and the whole $i$th isolated population should collapse.

To avoid such a tragedy, many times migration is known to come to rescue~\cite{SadhukhanPRR2021,SadhukhanPRE2021}. Hence, now we allow for migration between populations in the aforementioned model: Let us assume that the probability rate of migration of a cooperators and a defector from $i$th to $j$th population are $\mu^C_{ij}$ and $\mu^D_{ij}$ respectively. Thus, the rate of increase of cooperators in $i$th population due to immigration is  $\sum_{j\neq i}N_j^C\mu^C_{ji}$  and  the rate of decrease of cooperators in $i$th population due to emigration is $\sum_{j\neq i}N_i^C\mu^C_{ij}$. Similarly, the rate of increase of defectors in $i$th population due to immigration is  $\sum_{j\neq i}N_j^D\mu^D_{ji}$  and  the rate of decrease of defectors in $i$th population due to emigration is $\sum_{j\neq i}N_i^D\mu^D_{ij}$. Furthermore, we assume that each population communicates with every other $M-1$ populations and the migration rates are constant, i.e., they do not depend on the colonies from which they start to migrate and where they reach, i.e., $\mu^C_{ji}=\mu^C{/(M-1)}$ and $\mu^D_{ji}=\mu^D{/(M-1)}$ for all $i,j \in \{1,2,\cdots,M\}$. Finally, in the presence of migration, we rewrite Eq.~(\ref{evolution_Cooperator_without_migration}) and Eq.~(\ref{evolution_Defector_without_migration}) as follows:
\begin{subequations}
	\begin{eqnarray}
	\dot{N}_i^C&=&N_i^C\left[f_i^C(N_i^C,N_i^D)-\delta(N_i^C+N_i^D)\right]\nonumber\\ 
		& ~&-  \sum_{j\neq i}N_i^C\mu^C_{ij} +(1-h)\sum_{j\neq i}N_j^C\mu^C_{ji}, \label{evolution_Cooperator_with_migration}\\	
		\dot{N}_i^D&=&N_i^D\left[f_i^D(N_i^C,N_i^D)-\delta(N_i^C+N_i^D)\right]\nonumber\\ 
		& ~& -  \sum_{j\neq i}N_i^D\mu^D_{ij} +(1-h)\sum_{j\neq i}N_j^D\mu^D_{ji} \label{evolution_Defector_with_migration}.
\end{eqnarray}
\end{subequations}
\textcolor{black}{These are a set of $2M$ coupled nonlinear equations. Note that we have introduced a parameter, $h$, that is identically zero in the present case. Following paragraph makes the physical meaning of $h$ conspicuous.}

\textcolor{black}{Not every immigrant is welcomed in the new population. In the ants, every colony possesses a distinctive individuality due to which a  migrated ant is spotted, stopped from joining the new habitat~\cite{TSUJI1986,sanada2003encounter}, and eliminated. Due to this hostile behavior towards individuals different from their own habitat, not all of the migrated individuals survive. There is no known hostility towards emigration. Thus, the rate of increase of cooperators (defectors) in $i$th population due to immigration is  $p_s\sum_{j\neq i}N_j^C\mu^C_{ji}$ ($p_s\sum_{j\neq i}N_j^D\mu^D_{ji}$)  and  the rate of decrease of cooperators (defectors) in $i$th population due to emigration is unchanged---$\sum_{j\neq i}N_i^C\mu^C_{ij}$ ($\sum_{j\neq i}N_i^D\mu^D_{ij}$); here, $p_s$ is the probability of survival after the migration. Naturally, $h\equiv 1-p_s$, can be termed hostility: lower the survival rate, higher the hostility.}

\textcolor{black}{Obviously, Eq.~(\ref{evolution_Cooperator_with_migration}) and Eq.~(\ref{evolution_Defector_with_migration}) constitute an analytically intractable set of equations,} but it is amenable to some stability analysis. {\color{black} We have categorized the fixed points into two classes: homogeneous and non-homogeneous. A homogeneous fixed point corresponds to the metapopulation state where all the constituent colonies have same numbers of cooperators and defectors, i.e., mathematically, $N^C_i=N^C_j$ and $N^D_i=N^D_j$~$\forall i,j$; whereas, the fixed points which are not homogeneous are termed non-homogeneous fixed points.} The general stability analyses about all possible fixed points is however almost impossible because of the plethora of fixed points the system possess. Even with a metapopulation with two populations, it can have more than twenty fixed points. 
Nevertheless, homogeneous fixed points are limited and we can gain some insight by analyzing them. These fixed points correspond to the state of the metapopulation with the populations being in identical equilibrium states. Since our motivation is to elucidate the constructive effect of hostility  in metapopulation, we can succinctly accomplish that by working with the smallest possible metapopulation, i.e., $M=2$: While such a set-up loses the higher order intricacies that can appear due to the specific network topology of a metapopulation, the fundamental idea put forward in this paper should remain valid qualitatively even for $M>2$. Hence, henceforth, we stick with this choice.

{\color{black}
\section{Stability Analysis of Extinction State}
\label{app:A}
{\color{black}The stability analysis of the non-homogeneous fixed points is quite straightforward: We calculate the expression of the Jacobian matrix of the dynamical equations, Eq.~(\ref{evolution_Cooperator_with_migration}) and Eq.~(\ref{evolution_Defector_with_migration}), linearized about every fixed point and calculate its eigenvalues. If real parts of all the eigenvalues for the fixed point are negative, then the fixed point is stable; otherwise, it is unstable. Likewise the stability analysis of the homogeneous fixed points is also rather simple. However, it is interesting to note the non-triviality of stability analysis of the extinction state, $(N_1^C,N_1^D,N_2^C,N_2^D)=(0,0,0,0)$ (a homogeneous fixed point). The issue is that the ratio of the number of cooperators or defectors to the total population size is undefined ($0/0$). In fact, even to adjudge that $(0,0,0,0)$ is actually a fixed point, we need $x_i\equiv N_i^C/(N_i^C+N_i^D)$ to have a finite limiting value at the fixed point.}

Before we explain our adopted method of doing the stability analysis, it is probably instructive to study a simple toy model that is plagued by the similar problem. Consider a two dimensional autonomous flow:
\begin{subequations}
	\label{eq:toy}
	\begin{eqnarray}
		&&\dot{u}=-u-\frac{u^2}{v},\\
		&&\dot{v}=-v-\frac{v^2}{u},	
	\end{eqnarray}
\end{subequations}
where the phase space $u$--$v$ is $[0,\infty)\times[0,\infty)$.
It is obvious that the flow for any initial condition approaches $(u,v)=(0,0)$. However, it is a bit of a problem in terming this a fixed point in the light of the fact that $\dot{u}=0$ and $\dot{v}=0$ leads to algebraic equation with $0/0$ form at $(0,0)$. What comes to our rescue is the existence of the asymptotic value of $\gamma\equiv u/v$ (and $v/u$) as the system evolves. We find
\begin{eqnarray}
	\dot{\gamma}=1-\gamma^2,
\end{eqnarray}
whose physically allowed fixed point is $\gamma=u/v\to 1$ which is stable as well. Hence, we conclude that at $(u,v)=(0,0)$, $u/v$ takes the dynamics-driven limiting value that is unity. With this in mind, we can now safely, calculate $(0,0)$ as the fixed point of Eq.~(\ref{eq:toy}) by factoring $u^2/v$ as $\gamma u$. In fact, now even the linear stability analysis of  Eq.~(\ref{eq:toy}) about $(0,0)$ can be carried out because all the terms (in the Jacobian) in the form of $\gamma^2$ (or $1/\gamma^2$) calculated at the fixed point are not indeterminate but finite.

Now, coming back to the issue of dealing with the stability analysis about the fixed point $(0,0,0,0)$, it is of convenience to rewrite Eq.~(\ref{evolution_Cooperator_with_migration}) and Eq.~(\ref{evolution_Defector_with_migration}) in terms of the total number of individuals, $N_i\equiv N_i^C+N_i^D$ and the frequency of cooperators, $x_i=N_i^C/N_i$ as follows:
\begin{subequations}
	\begin{eqnarray}
		{\dot{x}_i}&=&x_i(1-x_i)\left[f_i^C-f_i^D+\mu^D-\mu^C \right] \nonumber\\
		&+&(1-h)\sum_{j \neq i} \frac{N_j}{N_i}\left[\mu^C x_j(1-x_i)-\mu^Dx_i(1-x_j)\right],\qquad\,\,\,\label{evolution_frequency_with_migration}\\
		{\dot{N}_i}&=&N_i\left[x_i\left(f_i^C-\mu^C\right)+(1-x_i)\left(f_i^D-\mu^D\right)\right]	\nonumber\\
		&+ &(1-h)\sum_{j \neq i}N_j\left[x_j\mu^C+(1-x_j)\mu^D\right]-\delta N_i^2, 
		\quad\label{evolution_total_with_migration}
	\end{eqnarray}
\end{subequations}
\textcolor{black}{where we have set $M=2$}.
In these notations, the extinction state corresponds to $N_i=0$ $\forall i$ irrespective of what value $x_i$'s take.

We note the factor $N_j/N_i$ in Eq.~(\ref{evolution_frequency_with_migration}) which is $0/0$ at the extinction state. We hope to associate a limiting value to this factor so that we can find fixed points of Eq.~(\ref{evolution_frequency_with_migration}). To this end, we write dynamical equation for $\gamma_{ij}\equiv N_i/N_j$ $\forall i\ne j$:
\begin{eqnarray}
	\dot{\gamma}_{ij}&=&\gamma_{ij}\left[x_i(f_i^C-\mu^C) +(1-x_i)(f_i^D-\mu^D) \right]\nonumber \\
	&&-\gamma_{ij}\left[x_j(f_j^C-\mu^C) +(1-x_j)(f_j^D-\mu^D)\right]  \nonumber\\
	&& +(1-h)\sum_{k\neq i}\gamma_{kj}\left[x_k\mu^C+(1-x_k)\mu^D\right]\nonumber\\
	&& -(1-h)\gamma_{ij}\sum_{k\neq j}\gamma_{kj}\left[x_k\mu^C+(1-x_k)\mu^D\right]\nonumber\\
	&&+\delta N_i (1-\gamma_{ij}).
	\label{eq:gamma}
\end{eqnarray}
It is clear that if $N_i=0$ (and arbitrary fixed $x_i$), Eq.~(\ref{eq:gamma}) possess at least one finite value of $\gamma_{ij}$ as its fixed point; specifically, if $N_i=0$ and $x_i=0$, ${\hat {x}}$ or $1$ (i.e., the homogenous fixed points of Eq.~(\ref{evolution_frequency_with_migration}) with fixed finite $\gamma_{ij}$), $\gamma_{ij}=1$ is the stable fixed point. Recall that ${\hat {x}}\equiv ({-|S|+\nu})/(T-1-|S|)$. Of course, there could be situations where $x_1\ne x_2$ and $N_1=N_2=0$ along with finite $\gamma_{ij}\ne 1$ correspond to an extinction state; however, these situations are best tackled numerically as we discuss later. One should not be surprised that the extinction state in the $(x_i,N_i)$ coordinates correspond to some specific values of $x_i$'s (and not arbitrary ones) because the limiting values of $\gamma_{ij}$'s decide how the extinction is approached.

The stability of extinction state as marked by the fixed points of Eq.~(\ref{evolution_frequency_with_migration}) and Eq.~(\ref{evolution_total_with_migration}), with $\gamma_{ij}$ having a limiting value,  can be found out by doing linear stability analysis about the homogeneus fixed points $(x_i^*,N_i^*)=(0,0),\,(1,0)$~and~$({\hat {x}},0)$ $\forall i$ and also about the non-homogeneous ones where $N_i=0$ but $x_1\ne x_2$ ($x_i\ne0$). The former is analytically tractable but the latter is accessible only numerically.
\begin{figure}
	\centering
	\includegraphics[width=0.8\linewidth]{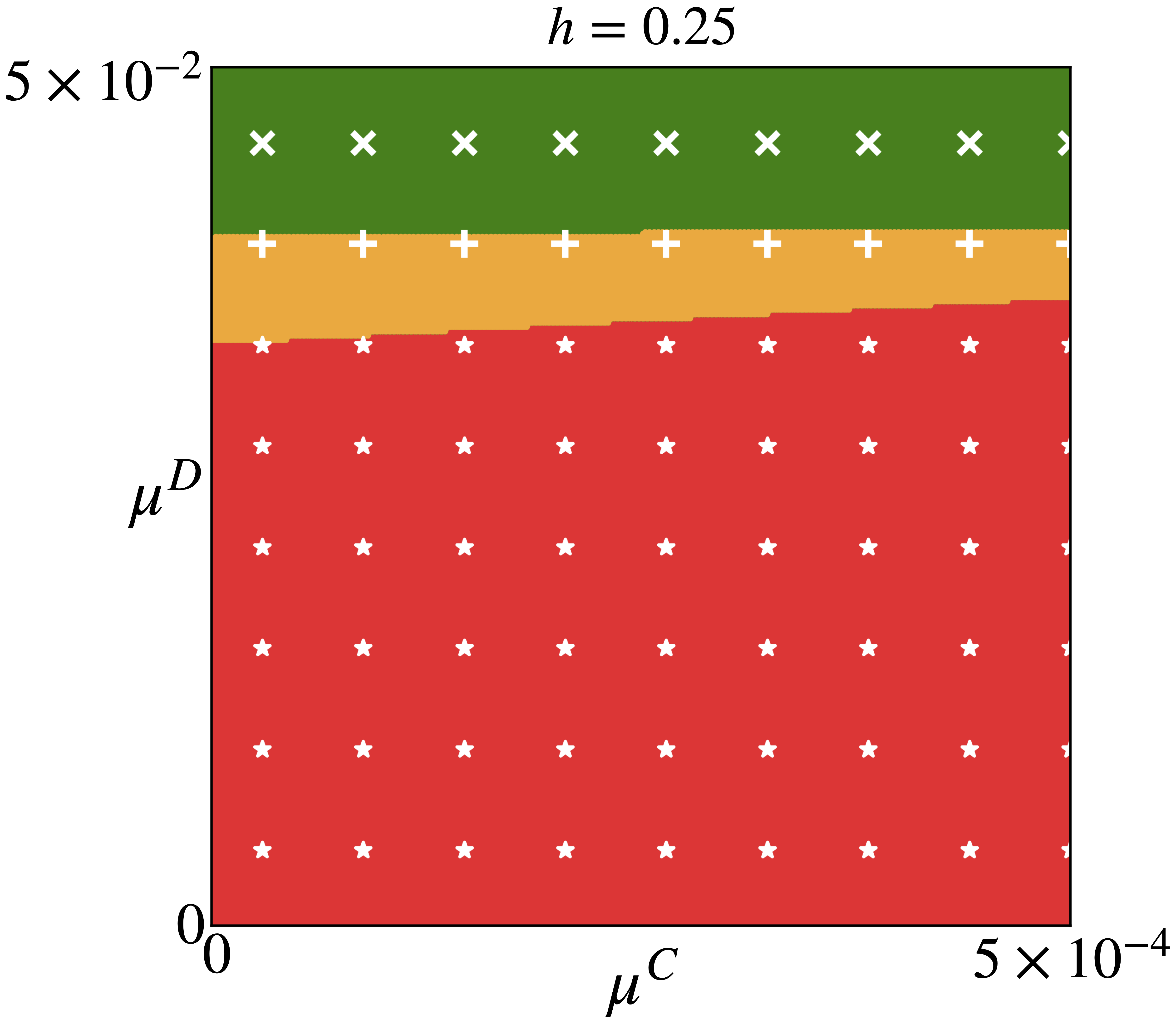}\\
	\caption{{\color{black}Numerics validate the stability diagram: We illustrate the validation of the analytically obtained stability diagram in Fig.~\ref{fig1_migration}(d). Extinction, bistability and heterogeneous existence of the metapopulation are respectively marked by the white markers $*$, $+$ and $\times$ which represent the direct numerical time evolution of  Eq.~(\ref{evolution_Cooperator_with_migration}) and Eq.~(\ref{evolution_Defector_with_migration}). For the numerics, we uniformly divide $\mu^C$--$\mu^D$ plane  in  $8 \times 8$ grid. At each grid point, we evolve ten randomly chosen initial conditions (for a time $t=15000$ by the fourth order Runge--Kutta method with discrete time step $dt=10^{-3}$) whose final states match with the corresponding analytical results: The red, the yellow, and the green patches---which respectively indicate extinction, bistability and heterogeneous existence---exclusively contain $*$, $+$ and $\times$ markers respectively.}}
	\label{fig_3_validation}
\end{figure}

Straightforward calculations reveal that $(1,0)$ is always unstable, whereas $(0,0)$ is stable when $|S|>\nu$, where $\nu\equiv h(\mu_D-\mu_C)$. This means that in the $\mu^C$--$\mu^D$ space, the curve dividing the regions of extinction and survival of the metapopulation is given by $\mu^D=\mu^C+|S|/h$: The region below this curve denotes stable extinction state. Consequently, as the hostility increases, the possibility of extinction decreases. In principle, such dividing curves should exist for other representations (where $x_i\ne0$) of the extinction state, although writing their explicit expressions is very cumbersome. However, what we can prove is that all such curves lie below a certain curve (henceforth referred as `existence boundary curve') that serves as the boundary below which non-extinction states do not exist. To this end, let us rewrite Eq.~(\ref{evolution_total_with_migration}) as follows:
\begin{eqnarray}
	\dot{N}_i=R_i(x_i,x_j, \gamma_{ji})N_i\left[1-\frac{\delta N_i}{R_i(x_i,x_j, \gamma_{ji})}\right],
\end{eqnarray}
where
\begin{eqnarray}
	R_i(x_i,x_j, \gamma_{ji})&\equiv& \left[x_i\left(f_i^C-\mu^C\right)+(1-x_i)\left(f_i^D-\mu^D\right)\right]	\nonumber\\
	&+& (1-h)\sum_{j \neq i}\gamma_{ji}\left[x_j\mu^C+(1-x_j)\mu^D\right].\quad\,
\end{eqnarray}
Now we note that at any fixed point with $x_1\ne x_2$ ($x_i\ne0$), the non-extinction state, when it exists, is $N_i=R_i(x_i,x_j, \gamma_{ji})/\delta$ which must be a positive quantity that in turn means that $N_i=0$ (i.e., the extinction state) is unstable. We should emphasize here  that the homogeneous extinction state represented by $(x_i=0,N_i=0)$, however, can stably coexist with the non-extinction state because in that case $ R_i(x_i,x_j, \gamma_{ji})=-\mu^D(1+\gamma_{ji})$ is always negative. This mathematically is the source of bistability in our system: Extinction and non-extinction states can stably co-exist---which one is realized depends on the specific initial composition of the metapopulation. In Fig.~(\ref{fig_3_validation}), we illustrate a validation of the fact that the phase diagram in $\mu^C$--$\mu^D$ space found analytically using above arguments should match with the results found using direct numerical evolution of the corresponding system.}

\section{Results}
\subsection{\textcolor{black}{Without hostility ($h=0$)}}
There exist two possible homogeneous fixed points, $({N}_i^{C^*},{N}_i^{D^*})$, of Eq.~(\ref{evolution_Cooperator_with_migration}) and Eq.~(\ref{evolution_Defector_with_migration})---\textcolor{black}{$h$ set to zero---}inside the physically allowed region of the phase space: $(0,0)$ and $(1/\delta,0)$ $\forall i$. The first fixed point corresponds to extinction of the metapopulation, while the second one corresponds to  the sustenance of the metapopulation at its most prosperous possible state exclusively composed of cooperators. They are stable if $|S|>0$ (see Sec.~\ref{app:A}) and $T-1<0$ respectively. 

 Furthermore, there are five independent parameter in our system. Their values are dependent on specific details of the system in question. As our main parameters of interest are the migration rates and hostility, it helps to fixed $S$, $T$, and $\delta$ to keep the discussion {focused}. To this end, it is helpful to use the values that are appropriate in the ant colony just for concreteness.

First, we take $\delta=10^{-4}$ so that the carrying capacity in our model becomes $10^4$, since the observed colony size of {\it Pristomyrmex punctatus} is of the order of $10^{4}$~\cite{Tsuji1995}. Naturally, we should have payoff of a defector interacting with a defector {\color{black}to} be zero. Since the per capita brood production is of the order one in the absence of any defectors, the payoff of a cooperator interacting with cooperator may be taken as one. Note that the diagonal elements of the general payoff matrix as introduced earlier are $1$ and $0$ anyway. The defectors definitely get more payoff that than cooperators when they interact with cooperators. Hence we take $T=1.5>1$. However, a cooperator in a population exclusively composed of defectors produces no brood~\cite{Dobata2013} but on top of it pays cost by working for the colony; hence, the corresponding payoff---$S$ in our notation---should be negative, that we take as $-0.01$. As long as the ordinal relationship between the payoff elements is maintained, one expects the results to be qualitatively same. The numbers we have specifically chosen closely match with the ant population to the best of our inferences from the existing literature on the ants. Finally, the migration rates per generation of cooperators and defectors are know to be of the order of $10^{-5}$ and $10^{-2}$ respectively~\cite{Dobata2013}. However, we let $\mu^C$ and $\mu^D$ vary over a larger range to further theoretical predictions.

For the aforementioned parameter values, one quick observation (Fig.~\ref{fig1_migration}a) is that all the {homogeneous} fixed points, {except} the one that corresponds to the extinction, are always unstable whether the migration is asymmetric or not. However, for high enough defector migration rate, a stable internal non-homogeneous fixed point appears. Of course, due to indistinguishability (as far as the parameter-values are considered) of the two colonies, any non-homogeneous fixed point must appear in pair. Depending on the precise initial condition, symmetry breaking occurs and one in the stable pair is the observed equilibrium solution. In summary, we have a multistable system where the extinction is avoided if cooperator fraction is above a threshold and the migration is asymmetric. The surviving state is, however, not {homogeneous} and, more importantly, it requires quite high value of $\mu^D$ ($\gg10^2$, the value for the ants) for its existence. This motivates us to bring in the phenomenon of hostility.
\subsection{\textcolor{black}{With hostility ($h\ne0$)}}
\textcolor{black}{As discussed earlier, since not every immigrant is allowed into a new population, the ants may be deemed hostile. The degree of hostility is measured by  parameter $h$ that is the probability of migrant's death in the new colony. Hence, in the presence of hostility, we consider [Eq.~(\ref{evolution_Cooperator_with_migration}) and Eq.~(\ref{evolution_Defector_with_migration})] but now with $h\in(0,1]$.}

Once again the landscape of the homogeneous fixed points, $({N}_i^{C^*},{N}_i^{D^*})$, is of interest. Here, the earlier (when $h=0$) two possible homogeneous corner fixed points---$(0,0)$ and $(1/\delta,0)$---{modify} to $(0,0)$ and $([{1-\mu^Ch}]/{\delta},0)$. The latter one (the prosperous state with full of cooperators) is stable if $T-1<\nu$, where $\nu\equiv h(\mu^D-\mu^C)$. If $|S|>\nu$, then the state of extinction, i.e., $(0,0)$ fixed point, is stable. Additionally, if either $|S|<\nu<T-1$ or $|S|>T-1>\nu$, then the extinct state may again be stable for some particular combination of the parameters which however has a cumbersome expression. Also, if either $|S|<\nu<T-1$ or $|S|>T-1>\nu$, a new homogeneous internal fixed point appears such that $(N^{C^*}_i/N^{D^*}_i,N^{C^*}_i+N^{D^*}_i)=({\hat {x}},{\hat {N}})$ where ${\hat {x}}\equiv ({-|S|+\nu})/(T-1-|S|)$ and  ${\hat {N}}\equiv[{\hat {x}}\{{\hat {x}}+(1-{\hat {x}})(T-|S|)\}-h\{{\hat {x}}\mu^C+(1-{\hat {x}})\mu^D\}]/{\delta}$. When there is no hostility, this particular fixed point moves out of the physically allowed region of the phase space. The {non-homogeneous} fixed points are practically accessible only numerically.

Since hostility induces less survival of the migrated ants, it may appear that the metapopulation is more prone to extinction. Interestingly, this is not the case: In the presence of asymmetric migration, the internal {homogeneous} fixed point is stabilized in the metapopulation with hostile individuals and extinction is averted. {\color{black}Moreover, the population size increases with hostility for fixed migration rates~Fig.~\ref{fig_2_toc}(c).}  In fact, even a stable {non-homogeneous} state of the population appears (as was the case when $h=0$) but now at much lower values of $\mu^D$ ($<10^{-2}$) so that the survival of ant metapopulations may be attributed to the hostile behavior whose existence from natural selection perspective is, thus, justified. .

\begin{figure}
	\centering
	\includegraphics[width=1\linewidth]{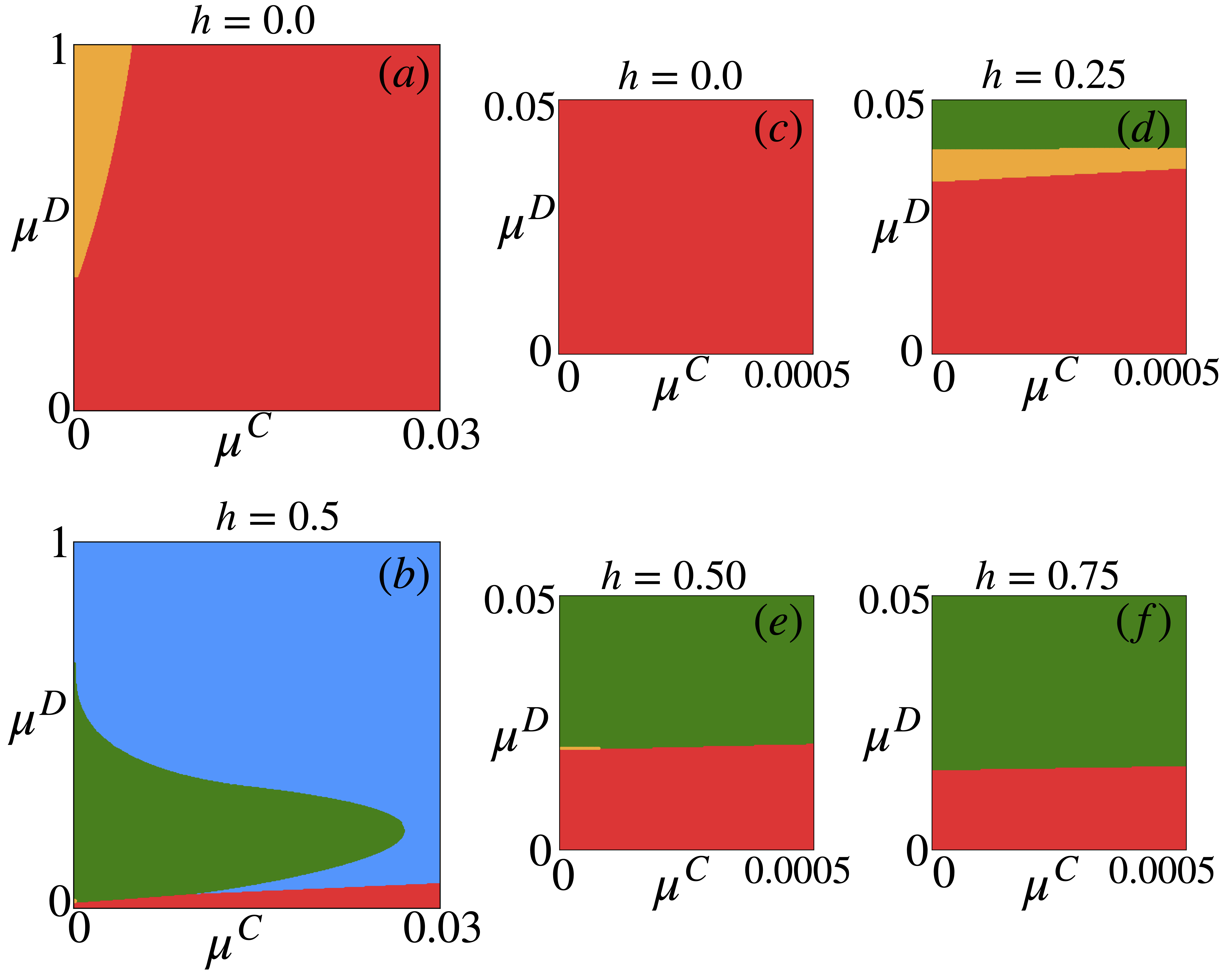}
	\caption{Hostility averts extinction: This figure depicts the stability diagram of  Eq.~(\ref{evolution_Cooperator_with_migration}) and Eq.~(\ref{evolution_Defector_with_migration}) with PD ($T=1.5$ and $S=-0.01$) as the underlying game structure. We also fix $\delta=10^{-4}$ and $M=2$. By keeping the hostility ($h$) fixed, we vary $\mu^C$  along $x$-axis and $\mu^D$ along $y$-axis for generating every subplot. The red, yellow, green, and blue colors respectively depict the parameter region for which the metapopulation goes extinct, sustains bistability between extinction and non-homogeneous extant state, sustains non-homogeneous extant state, and sustains homogeneous extant state. Subplot $(a)$ is for the non-hostile case while subplot $(b)$ is for the hostile case with $h=0.5$. The lower row of subplots $(c)$ to $(d)$ is zoomed in plots about the origin and the value of hostility increases from $(a)$ to $(d)$ as marked in the figure; here the ranges of $\mu^C$ and $\mu^D$ are close to what was reported for {\it Pristomyrmex punctatus}~\cite{Dobata2010,Dobata2013}.}
	\label{fig1_migration}
\end{figure}
{\color{black}
\subsection{Sustaining Non-homogeneous States}
\label{app:B}
The role of asymmetric migration should again be emphasized: In case, the migration is symmetric ($\mu^C=\mu^D$), we can show (elaborated below) that no amount of hostility can overcome extinction of the metapopulation (see Fig.~\ref{fig1_migration}).

First we observe that case of symmetric migration always leads to extinction irrespective of how high the hostility is. First we note that when $\mu^C=\mu^D$, the homogeneous fixed point $(x_i=0,N_i=0)$ is always stable (since $|S|>\nu=0$). It leaves the possibility of coexisting non-extinction state. However, even this is not a possibility as we see next. The non-extinction state must have $x_i\ne0$ since with $x_i=0$ only possible fixed point is $(x_i=0, N_i=0)$ $\forall i$. We recast Eq.~(\ref{evolution_frequency_with_migration}) as follows:
\begin{eqnarray}
	\dot{x}_i=x_i(1-x_i)\left(f_i^C-f_i^D\right)+(1-h)\gamma_{ji}(\mu^Cx_j-\mu^Dx_i),\qquad
	\label{symmetric_migration}
\end{eqnarray}
where $j\ne i$.
Within the paradigm of PD, $f_i^C-f_i^D<0$, and hence unless $\mu^Cx_j-\mu^Dx_i$ is positive, the system cannot evolve to nonzero values of $x_1$ and $x_2$. But for $\mu^C=\mu^D$, it leads to the contradicting conditions, respectively, $x_2>x_1$ and $x_1>x_2$. In conclusion, symmetric migration always leads to extinction of the metapopulation playing PD irrespective of initial compositions and parameter values.

Consequently. in a  discussion on the physical mechanism behind the emergence and sustenance of non-homogeneous states we have to restrict ourselves to the case of asymmetric migration. It is a straightforward observation from Eq.~(\ref{evolution_frequency_with_migration}) and Eq.~(\ref{evolution_total_with_migration}) that if the system evolves with the initial conditions, $x_i(0)=x_j(0)$ and $N_i(0)=N_j(0)$ $\forall i \ne j$, the resulting trajectories always remain on the hypersurface given by $x_i=x_j$ and $N_i=N_j$ which, thus, is an invariant manifold. In $N_1^C$--$N_1^D$--$N_2^C$--$N_2^D$ phase space coordinates, this hypersurface is expressed as $N_1^C=N_2^C$  and $N_1^D=N_2^D$. Therefore, the existence of the non-homogeneous state requires the initial condition must not follow these equalities. 

To understand the physical picture behind the emergence of non-homogeneous state, consider  an initial condition such that $N_1(0)^C>N_2^C(0)$  and $N_1^D(0)>N_2^D(0)$. With such an initial condition, we note from Eq.~(\ref{evolution_Cooperator_with_migration}) and Eq.~(\ref{evolution_Defector_with_migration}) that the individuals from population 1 migrate to population 2 at a larger number than the individuals who arrive at population 1 from the other. Obviously, the population 1 can be sustained only if  the defectors migrate at a higher rate than the cooperator (i.e., if $\mu^D>\mu^C$) such that the fitness benefit of the defectors is counter-balanced by the negative effect of the migration, i.e., $-\mu^D(N_1^D-N_2^D)<0$. Thus the population is maintain a constant frequency of cooperator and its total number of individuals do not decrease further.  In population 2, the frequency of the cooperators becomes very low because of higher number of immigrant defectors. However, it does not become  identically zero if $\mu^C \ne 0$:  In this case, population 1 acts as a source of cooperators for population 2, rendering the frequency, $x_2$,  bounded by $x_2>x_1\mu^C/[x_1\mu^C+(1-x_1)\mu^D]$. Of course, if $\mu^C=0$ then population 2 has no source of cooperators present and population 2 becomes full of defectors in the long run but it is not extinct. }

\subsection{Tragedy of the commons}
 In the literature~\cite{Rankin2007}, whether the TOC refers to complete or partial destruction of the resource (say, workforce) leads to the naming of the respective tragedy as collapsing or component. In any typical evolutionary system, the fact that significant levels of cooperation---in spite of the selfish interests of cheaters---are observed, begets the question of why a component TOC does not always deteriorate further into collapsing one, or why even component TOC is not present in certain cases (mostly where cooperation levels are very high). We have shown that territoriality induced hostility is one answer to this question.

\begin{figure*}
	\centering
	\includegraphics[width=1\linewidth]{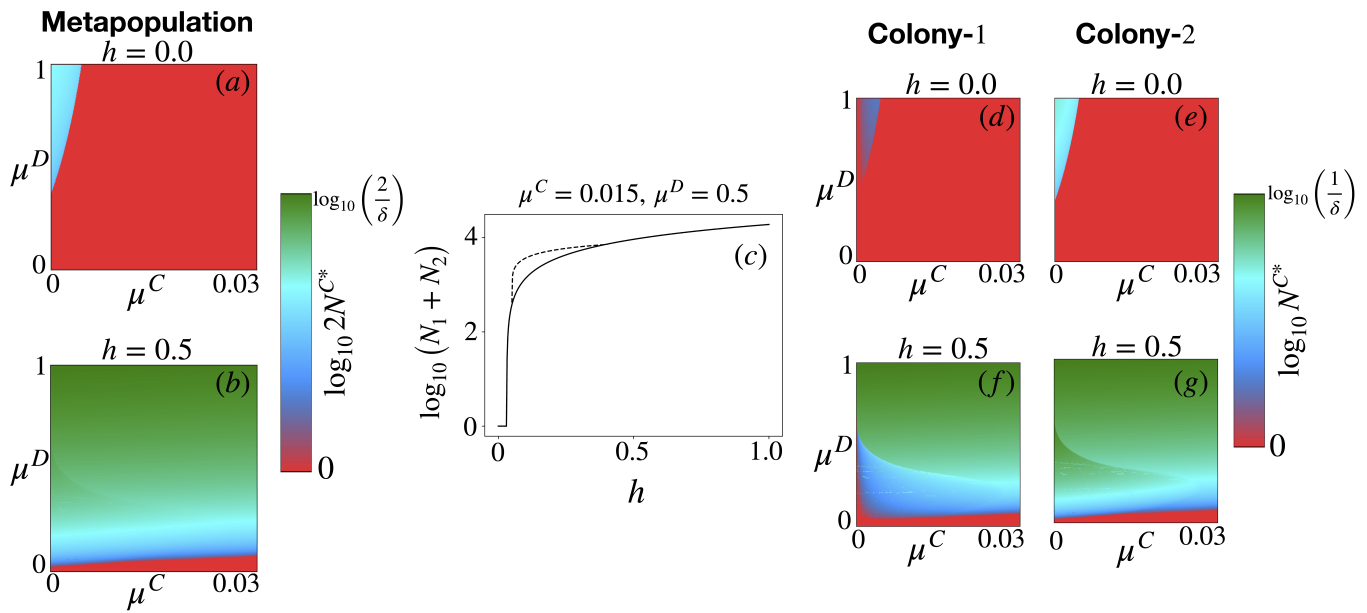}
	\caption{Hostility averts the  TOC:  The first row represents the realization and the prevention of  the TOC in  metapopulation, whereas the second row represents the same for the constituent populations/colonies separately. We find that the presentation is conspicuous if we plot the logarithm of the total number of cooperators; for the extinction state we replace $N_1^C=N_2^C=0$ with a small number, $10^{-20}$. Fixed parameters are same as in Fig.~\ref{fig1_migration}. The red and the darkest green regions respectively depict the complete realization and the prevention of TOC, whereas the all other colors represent the partial TOC.  We can see (in subplots $(a)$, $(d)$ and $(e)$) that the TOC is an unavoidable fate without hostility;  however, (as shown in subplots $(b)$, $(f)$ and $(g)$) with hostility $(h=0.5)$, the prevention of TOC always possible for suitable parameter values.
	{\color{black}Subfigure (c) shows the increase in size of metapopulation with the hostility: The solid line corresponds to homogeneous stable fixed point, while the dashed line represents the non-homogeneous fixed point. Here $\mu^C=0.015$ and $\mu^D=0.5$.}}
	\label{fig_2_toc}
\end{figure*}

The workforce---a social good formed by cooperation---in our set-up is simply measured by the number of cooperators. In the setting of PD, we note that in the absence of hostility, collapsing TOC is inevitable; except that sometimes certain migration rates that allow for bistability, can come to rescue. As hostility kicks in, the collapsing TOC can be either converted to a component one, or even be completely prevented at high enough defector migration rate~(Fig.~\ref{fig_2_toc}). 

{\color{black}In passing, we remark that an immediate criticism of the results presented till now could be about the usage of PD when in an ant colony, multiplayer interaction---and hence the framework of PGG---is more realistic: The workforce~\cite{Wenseleers2004,Dobata2012} in a eusocial insect colony is the social good that  may be envisaged as the product of a PGG being played by a group of individuals. Actually, the idea we want to convey is most simply said through the PD and the results obtained using PGG are qualitatively similar~(see Appendix~\ref{app:PGG}). Nevertheless, the reference of PGG brings the corresponding metapopulation's state under the purview of TOC.}

{\color{black}
	\section{Multi-colony Metapopulation}
	We have conclusively established that hostility, in the presence of asymmetric migration, averts TOC in the setting of two-colony metapopulation. It is easy to argue that this assertion remains qualitatively intact even in the case of multi-colony metapopulation ($M>2$) irrespective of the fact whether the metapopulation is structured or not. This is because if the hostility is same towards every colony, there is nothing in our model that distinguishes a nearby colony from a far-off colony; in fact, in our model every colony may see intruders from different colonies as the ones coming from a single effective colony. Nevertheless, the investigation of multi-colony metapopulation becomes extremely relevant in the context of dear-enemy and nasty-neighbor effects which can be mathematically captured via unequal hostilities in different pairs of colonies.

\begin{figure}
	\centering
	\includegraphics[width=1.0\linewidth]{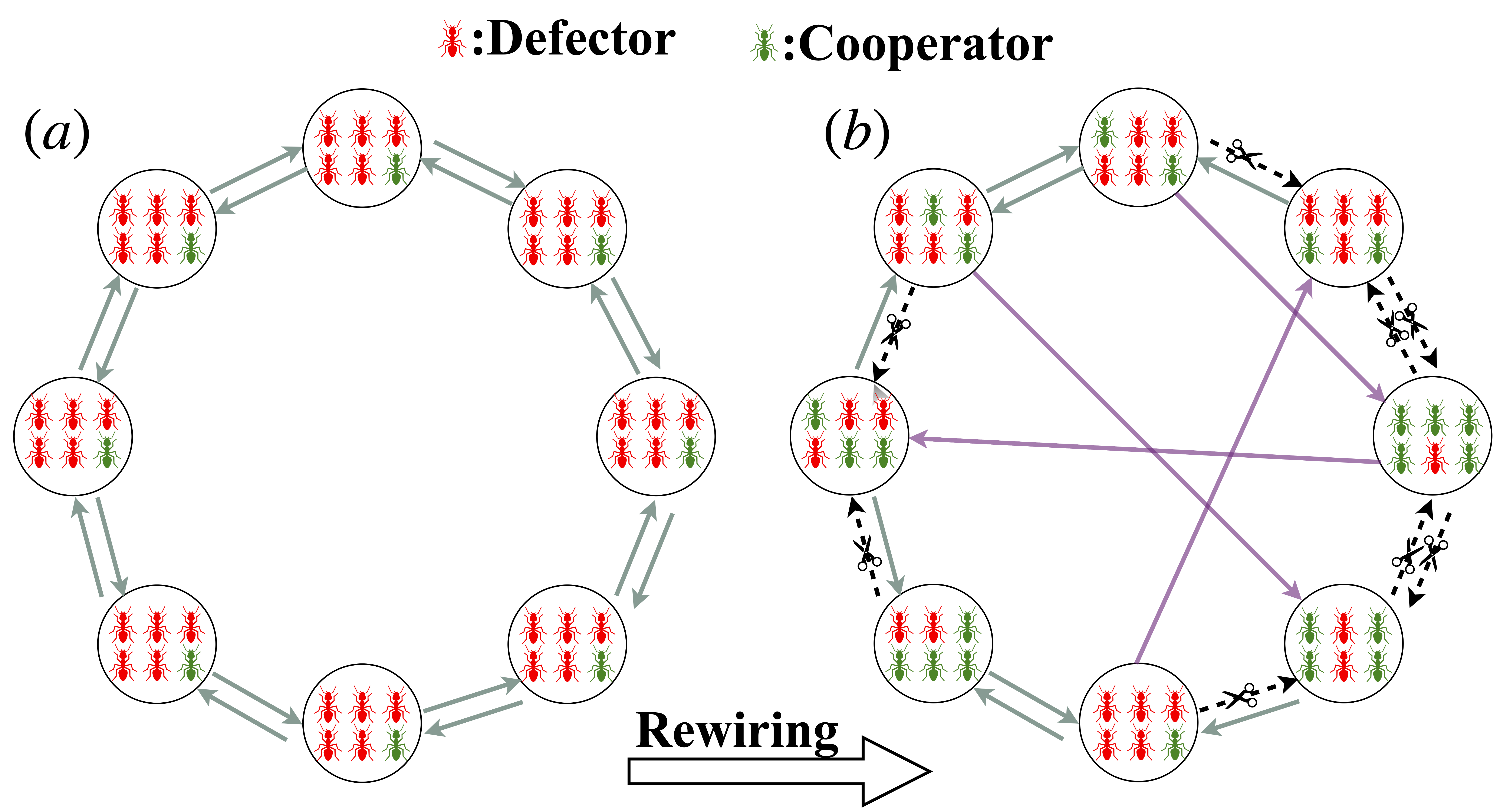}
	\caption{\color{black}Diagrammatic depiction of multi-colony metapopulation: Subfigure $(a)$ shows nearest neighbor interaction, and subfigure $(b)$ depicts the random rewiring at a time step. The directed arrow heads show the directed migrations from one colony to the other. A nearest-neighbor migration is indicated by solid green arrow, a broken connection (dashed arrow with scissors) indicates no nearest-neighbor migration at that time step and a new connection (violet arrow) indicates migration from far-off node.}
	\label{fig_6_meta_net}
\end{figure}
	
	Therefore, in order to mathematically address this, we consider a ring network (metapopulation) with $M$ nodes (colonies) as illustrated in Fig.~\ref{fig_6_meta_net}. To begin with there is only nearest neighbor interaction: The edges of the lattice are bidirectional and represent migration. However, intruders from far-off colonies can also come to a given colony. At any instant, thus, there is a probability ($p$, say) that an intruder is from far-off colony; naturally, $1-p$ is the probability that the intruder is from nearby colony. We incorporate this phenomenon by allowing random rewiring~\cite{2002_Sinha_PRE, SadhukhanPRR2021,SadhukhanPRE2021} in the network: The ring network dynamically evolves such that at every instant an edge between two adjacent colony may be severed and in its place a new edge with a randomly chosen (with probability $p$) far-off node is created. 
	
   Thus, if the hostilities towards the distant and the nearest neighbors are $h_d$ and $h_n$, respectively, then  Eq.~(\ref{evolution_Cooperator_with_migration}) and (\ref{evolution_Defector_with_migration}) modify to following mean-field equations for the ring network:
\begin{subequations}
	\begin{eqnarray}
		\dot{N}_i^C&=&N_i^C\left[f_i^C(N_i^C,N_i^D)-\delta(N_i^C+N_i^D)-\mu^C\right] +\nonumber\\ 
		& ~& (1-h_d)pN_\xi^C\mu^C_{\xi i}+(1-h_n)(1-p)\sum_{\langle ji\rangle }N_j^C\mu^C_{ji},\qquad\,\, \label{evolution_Cooperator_with_migration_hotility}\\	
		\dot{N}_i^D&=&N_i^D\left[f_i^D(N_i^C,N_i^D)-\delta(N_i^C+N_i^D)-\mu^D\right] +\nonumber\\ 
		& ~& (1-h_d)pN_\xi^D\mu^D_{\xi i}+(1-h_n)(1-p)\sum_{\langle ji\rangle }N_j^D\mu^D_{ji} \label{evolution_Defector_with_migration_rewire_hotility}.\qquad\,\,
	\end{eqnarray}
\end{subequations}
Here, $i$ is the focal node or colony and can take any value from $1$ to $M$. Due to the periodic structure of the network, $i$ and $i+M$ denote the same node. Subscript $\xi$ (not equal to $i-1$, $i$ or $i+1$) denotes the randomly chosen distant node. $\mu^C\equiv\sum_{j\ne i}\mu^C_{ij}$ and $\mu^D\equiv\sum_{j\ne i}\mu^D_{ij}$ are the total cooperators'/defectors' emigration rates from colony $i$. The cases $h_d<h_n$ and $h_d>h_n$, by definition, correspond to the nasty-neighbor effect and the dear-enemy effect, respectively.

For the case of {\it Pristomyrmex punctatus}, it is known~\cite{sanada2003encounter} that these ants show nasty-neighbor effect, i.e., $h_d<h_n$. Numerically, we observe (as seen in Fig.~\ref{fig_7_dear_enemy}) that with the nasty-neighbor effect metapopulation can avert the tragic extinction at lower defectors' migration rate. In Fig.~\ref{fig_7_dear_enemy}, for illustrative purpose, we use $M=5$, $p=0.1$ (assuming immigration is lower from far-off colonies) and divide $\mu^C$--$\mu^D$ plane  in  $48 \times 48$ grid. At each grid point, we evolve ten randomly chosen initial conditions (for a time $t=15000$ by the RK4 method with discrete time step $dt=10^{-3}$) whose final states are the red, the yellow, and the green patches, respectively, indicating extinction, bistability and heterogeneous existence.

This positive effect of nasty-neighbor may be comprehended by considering the average hostility, $ph_d+(1-p)h_n$: Since the probability of immigration from distant colonies is expected to be lower than that corresponding to the neighboring colonies, i.e., $p<1-p$, the average hostility is clearly higher for the case $h_d<h_n$ {\st{compared}compare} to the case $h_d>h_n$. Thus, given the observational data we have used for the ant in our numerics, the existence of nasty-neighbor effect in the ant is compatible with the evolutionary game-theoretic model presented in this paper.

\begin{figure}
	\centering
	\includegraphics[width=1.0\linewidth]{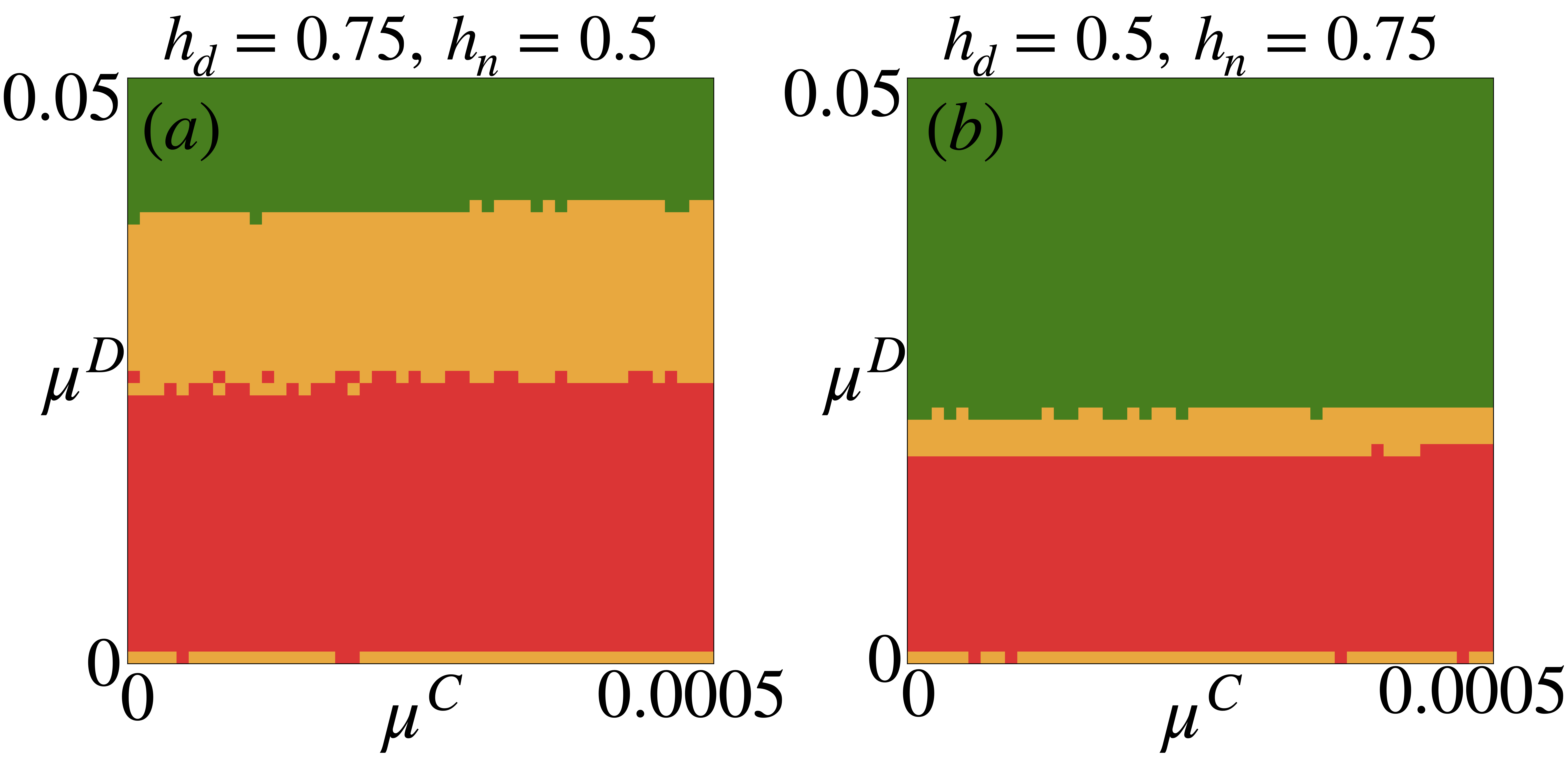}
	\caption{\color{black}The propensity of preventing the extinction is higher with the nasty-neighbor effect than with the dear-enemy effect. Subplots $(a)$ and $(b)$ are the stability diagram for the dear-enemy ($h_d=0.75>0.5=h_n$) and nasty-neighbor ($h_n=0.75>0.5=h_d$) effects, respectively. Rewiring probability has been fixed to $0.1$. Colors bears the same meaning and all other fixed parameters' values are same as in Fig.~\ref{fig1_migration}.}
	\label{fig_7_dear_enemy}
\end{figure}}

\section{Conclusion}

{\color{black}Why hostility should give rise to cooperation across metapopulation, which thus is prevented from going extinct, in the presence of asymmetric migration is intuitively quite simple: Defectors try to migrate to another colony more frequently than the cooperators but the hostile residents of the destination colony kill them; consequently, the total defector-fraction is kept in check. Our mathematical model very vividly captures this intuition. What is even more interesting is that---like any useful good model---there are some quite non-intuitive conclusions gathered through study of our model: Firstly, even in the absence of any hostility, asymmetric migration can lead to sustenance of cooperation if the colonies are of unequal sizes and with unequal cooperator fractions. In fact, we see in the bottom row of Fig.~(\ref{fig_2_toc}) that for very small migration rate of cooperators, the TOC is averted more effectively in one population than the other which is sustained even with high defector fraction~(see Sec~\ref{app:B}). (It is not uncommon to find an ant colony with quite high defector fraction up to almost $50\%$~\cite{Itow1984}.) Secondly, for relatively lower migration rates, hostility averts TOC and leads to non-homogeneous distribution of individuals across  colonies. Thirdly, equal sized stable colonies (homogeneous states) exist only when hostility is in action and such states do not coexist with extinction states---meaning they are much more robust survival states compared to the non-homogeneous states. Therefore, although there are various deciding factors behind unequal-size distribution in a metapopulation---something generically seen in nature (e.g., in {\it Pristomyrmex punctatus} ants~\cite{Tsuji1995})---it appears that hostility is one of them.

Furthermore, we would {emphasize} that this is the first theoretical work which has successfully explained the survivability of the ant species {\it Pristomyrmex punctatus}, in the face of defectors' selfish behavior, by using the direct observational data available on literature. Our mathematical model also shows that their ability to discriminate strangers from neighbors (with whom they interact multiple times) and to show more aggregation towards the latter is evolutionarily justified as this behavior further suppresses the possibility of extinction owing to selfishness of the defectors.}

Before we conclude, we must reiterate that in this paper, we have considered this ant society---for which we fortunately have access to some observational data in the literature---as a convenient representation of the similar systems across the living world and hence all the aforementioned lessons gained are, in general, qualitatively applicable to many other metapopulations as well.

While it is fascinating how much insight one has gathered using deterministic mean-field kind of dynamics and how well the dynamics conforms to the data available on {\it Pristomyrmex punctatus} ant, carefully investigating the system as a stochastic birth-death process~\cite{traulsen2005prl,lin2019prl,2021MCChaos,DasBairagya2023} is necessary, especially when the population size is not very high and the stochastic effects come to the fore. In fact, near the extinction state, the stochastic effects are at their peak; moreover, for very low values of $N_i$'s even talking about strategic interactions among group of large size is not meaningful. Hence, what we recognize as the extinction state in the deterministic model essentially a state with low enough population size such that it is almost an extinction but high enough population size that mean-field model still makes some sense. The deterministic model correctly tells whether an extinction state is approached of not; the precise dynamics about the such a low-populated state is, however, best studied stochastically. We would like to bring the effects of stochasticity, delays~\cite{Alboszta2004,Mittal2020}, structured matapopulation~\cite{Castle2020} into the scope of future investigations. Spatially restricted migration in structured population is known to avert extinction~\cite{Dobata2010}.  Also, extending this work to the cases with more than two kinds of individuals~\cite{Venkateswaran2019}---not just cooperators and defectors---is an exciting avenue of research as it opens up possibilities of more complex scenarios encountered in the multi-strategy games~\cite{Hajihashemi2022}.

\acknowledgements
The authors are indebted to Debashish Chowdhury for introducing them to the exciting research-works on the ants and {for} many related discussions. JDB has been supported by Prime Minister's Research fellowship (govt. of India) and SC acknowledges the support from SERB (DST, govt. of India) through project no. MTR/2021/000119.

\appendix

\section{PGG in Metapopulation}
\label{app:PGG}

Let us now discuss the case where one goes beyond the simultaneous two-player interaction to simultaneous multiplayer interaction, or in other words, we consider NPD ($N$-player PD) instead of PD. If we further consider the form of PD that corresponds to the additive model studied by Hamilton~\cite{HAMILTON_1964} and Trivers~\cite{1971_Trivers}, then it is equivalent to a PGG game~\cite{Hauert2003}. Specifically, the form of  the payoff matrix under consideration is
\begin{tiny}$\left[
\begin{array}{cc}
  b-c& -c   \\
  b& 0   
\end{array}
\right]$
\end{tiny}, where $b$ is the benefit rendered by a cooperator while incurring a cost $c$; defectors neither provide any benefit nor incur any cost. We note that the payoff matrix chosen in the main text is only ordinally equivalent to the aforementioned payoff matrix. In the setting of the PGG, suppose that a finite number of individuals in a population randomly form a group. All the cooperators  contribute towards a public good  which is utilized equally by the all members of the group; none of the defectors  contribute anything.
	
 We assume that the group is formed  within a population through binomial sampling where the probabilities of choosing a cooperator and defector respectively are the overall frequencies of the cooperators and the defectors in that population.  Thus, the probability of a focal cooperator finds herself in a group of size $G$ consisted of  $k$ other cooperators is $\binom{G-1}{k} x_i^k(1-x_i)^{G-1-k}$ $\forall k \in \{0,1,\cdots,G-1\}$, where $x_i$ is the frequency of the cooperators in the $i$-th population~\cite{HAUERT2006195, Patra2022}. Here,  $\binom{a}{b} $ is the binomial coefficient---`$a$ choose $b$'. Furthermore, we consider the public good is linearly proportional to the total contribution made by the cooperators. Accordingly, the effective fitness of a cooperator becomes 
	\begin{eqnarray}
		f^C&=&\sum_{k=0}^{G-1} \binom{G-1}{k}\left[\frac{\alpha c (k+1)}{G}-c\right] x_i^k(1-x_i)^{G-1-k} , \nonumber\\
		&=&\frac{\alpha c (G-1)x_i}{G} +\frac{\alpha c}{G}-c ,\label{cooperator_fitness_pgg}
	\end{eqnarray}
	where $c$ is the contribution by a cooperator and $\alpha$ is the synergy factor. Similarly, 
	\begin{eqnarray}
		f^D&=&\sum_{k=0}^{G-1} \binom{G-1}{k}\left[\frac{\alpha c k}{G}\right] x_i^k(1-x_i)^{G-1-k}\nonumber \\
		&=&\frac{\alpha c(G-1)x_i}{G} \label{defector_fitness_pgg}
	\end{eqnarray}
is the fitness of a defector.

\begin{figure}
	\centering
	\includegraphics[width=1.0\linewidth]{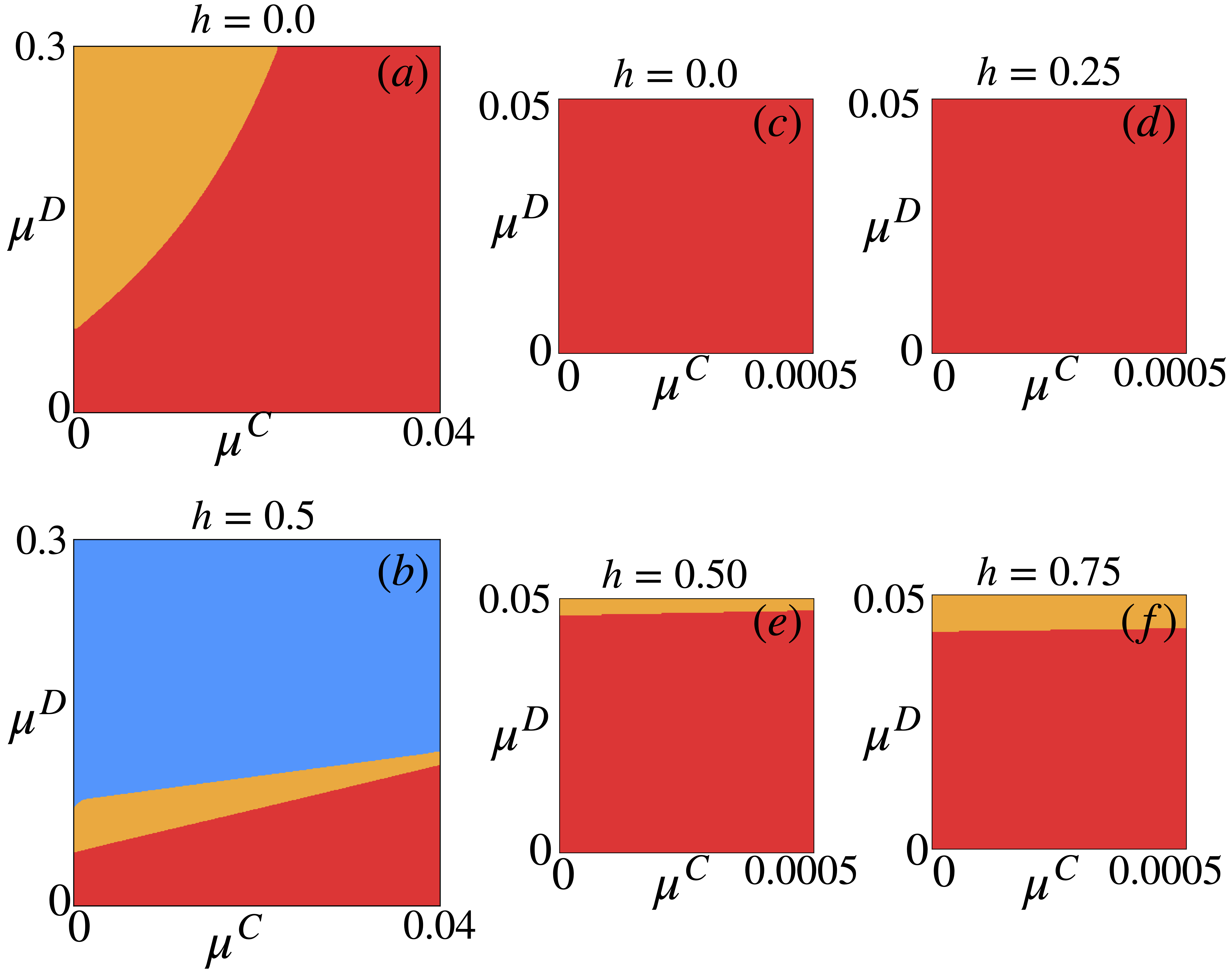}\\
	\caption{ Hostility averts extinction: This figure depicts the stability diagram of  Eq.~(\ref{evolution_Cooperator_with_migration}) and Eq.~(\ref{evolution_Defector_with_migration}) with PGG ($\alpha=5$, $G=6$ and $c=1/4$) as the underlying game structure. We also fix $\delta=10^{-4}$ and $M=2$. By keeping the hostility ($h$) fixed, we vary $\mu^C$  along $x$-axis and $\mu^D$ along $y$-axis for generating every subplot. The red, yellow, and blue colours respectively depict the parameter region for which the metapopulation goes extinct, sustains bistability between extinction and non-homogeneous extant state, and sustains homogeneous extant state. Subplot (a) is for the non-hostile case while subplot (b) is for the hostile case with $h=0.5$. The lower row of subplots (c) to (d) is zoomed in plots about the origin and the value of hostility increases from (a) to (d) as marked in the figure; here the ranges of $\mu^C$ and $\mu^D$ are close to what was reported for {\it Pristomyrmex punctatus}~\cite{Dobata2010,Dobata2013}.}
	\label{fig4_pgg_sd}
\end{figure}
\begin{figure*}
	\centering
	\includegraphics[width=1.0\linewidth]{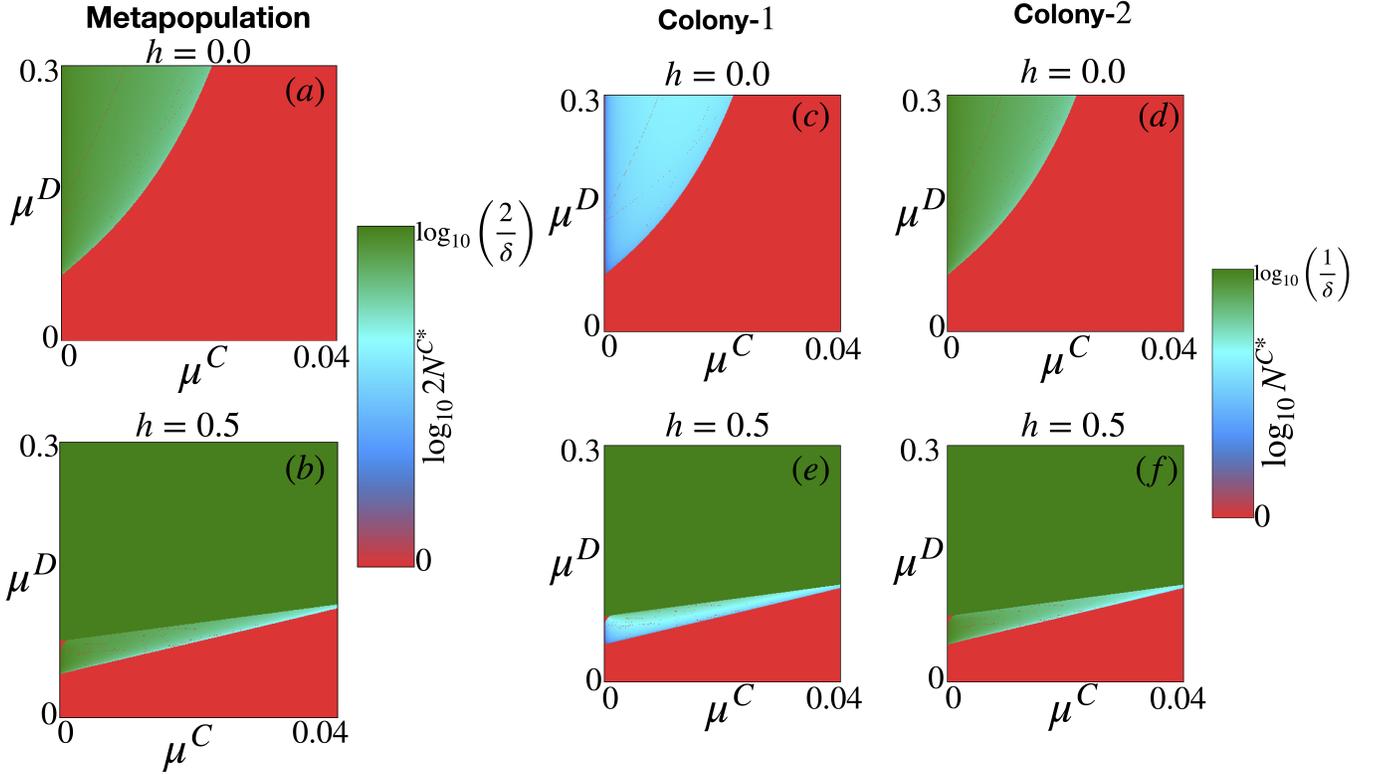}\\
	\caption{Hostility averts the  TOC:  The first row represents the realization and the prevention of  the TOC in  metapopulation playing PGG, whereas the second row represents the same for the constituent populations/colonies separately. We find that the presentation is conspicuous if we plot the logarithm of the total number of cooperators; for the extinction state we replace $N_1^C=N_2^C=0$ with a small number, $10^{-20}$. Fixed parameters are same as in Fig.~\ref{fig4_pgg_sd}. The red and the darkest green regions respectively depict the complete realization and the prevention of TOC, whereas the all other colours represent the partial TOC.  We can see (in subplots $(a)$, $(c)$ and $(d)$) that the TOC is an unavoidable fate without hostility;  however, (as shown in subplots $(b)$, $(e)$ and $(f)$) with hostility $(h=0.5)$, the prevention of TOC always possible for suitable parameter values.}
	\label{fig5_pgg_toc}
\end{figure*}

From the expression of these fitnesses, it is clear that the fitness of  a cooperator is always less than that of the defector if $\alpha/G<1$. This scenario leads the metapopulation towards the state of  no cooperators and eventually it collapses. Similar to the PD, here also we elucidate how the asymmetric migration and hostility prevents this tragic fate. Now with these fitnesses [Eq.~(\ref{cooperator_fitness_pgg}) and Eq.~(\ref{defector_fitness_pgg})],  Eq.~(\ref{evolution_Cooperator_with_migration}) and Eq.~(\ref{evolution_Defector_with_migration}) have only two homogeneous fixed  points, $({N}_i^{C^*},{N}_i^{D^*})$, inside the allowed region of the phase space: $(0,0)$ and $\left[c(\alpha -1)/\delta,0 \right]$.  The first fixed point indicates the extinction and the second one is the most prosperous state. The latter one is always unstable and the first one is always stable. Along with these two corner fixed points, there may exist some non-homogeneous internal fixed points whose existence and stability conditions  are only numerically accessible. Owing to indistinguishability (as far as the parameter-values are considered) of the two colonies, any non-homogeneous fixed point appears in pair. The stable pair consists of larger number of total cooperators compared to the unstable pair. Depending on the exact initial condition, degeneracy is lifted and one in the stable pair is the observed equilibrium solution. Thus, we have a multistable system where the extinction is bypassed if the migration is asymmetric and the number of cooperators is above a threshold. The surviving state is not homogenous and requires quite high value of $\mu^D$ ($\gg10^2$, the value for the ants) for its existence. 

Now if we introduce hostility, the two homogeneous corner fixed points--$(0,0)$ and $\left(c[\alpha -1]/\delta,0 \right)$--are modified to $(0,0)$ and $\left([c\{\alpha -1\}-h\mu^C]/\delta,0 \right)$. No homogeneous internal  fixed point exists.  Interestingly, in contrast with the case of without hostility, here the latter one (prosperous state) can become stable if $c(1-\alpha/G)<\nu$ (where $\nu=h(\mu^D-\mu^C)$); otherwise the first one (extinction state) is stable. The existence and stability conditions of the non-homogeneous fixed points are found numerically. One quick observation is that the bistability between the extinction state and the non-homogeneous fixed points is always present irrespective of the parameters' values, and always remains in the region where the prosperous state is unstable. However, with the hostility in action, the prosperous state can be stable for a lower value of $\mu^C \sim 10^{-5}$ and $\mu^D\sim 10^{-2}$ which is observed in the ant.The argument around Eq.~(\ref{symmetric_migration}) is also valid here meaning that the symmetric migration makes the extinction unavoidable fate for the metapopulation.

We present the summary of the results in Fig.~\ref{fig4_pgg_sd} and  Fig.~\ref{fig5_pgg_toc} where, for illustrative purpose, we use the parameters which closely encapsulates the fitnesses of the ant colonies ({\it Pristomyrmex punctatus}). The following values of the parameters, i.e., $c=1/4$ and $\alpha=5$, represent the order unity fitness of a cooperator in the absence of any defector~\cite{Tsuji1995}. We use the value of $G=6$~($>\alpha=5$) to exemplify the fitness of a defector that is more than of a cooperator~\cite{Dobata2013}. Finally, we let $\mu^C~\in~[0,0.04]$ and $\mu^D~\in~[0,0.3]$ to vary over a larger range to illuminate all possible outcomes. Hostility, $h$, can vary from $0$ to $1$.

{\color{black} One interesting observation is worth pointing out: While in the PD case, high enough hostility does not even let the extinction state coexist with the metapopulation state with $N_1,N_2>0$ and $N_1\ne N_2$ (see Fig.~\ref{fig1_migration}); in the case of the PGG (see Appendix~\ref{app:PGG}), hostility cannot do so (see Fig.~\ref{fig4_pgg_sd}). This coexistence between all defector state (that essentially leads to extinction) and a mixed state (mixture of cooperators and defectors) is reminiscent of coexistence of the PD and the anti-coordination game solutions---something that is achieved in the presence of the nonlinear PGG only~\cite{Patra2022}. It, thus, appears that migration is a probable route to such game-coexistences in the linear PGG.}

\bibliography{Bairagya}
\end{document}